\RequirePackage{graphicx}
\documentclass[aps,prl,preprint,groupedaddress,notitlepage]{revtex4-1}
\selectfont

\preprint{NuFact15 - Rio de Janeiro, Brazil - August 2015}

\begin{document}


\title{Experimental status of neutrino scattering}
\thanks{\it Presented at NuFact15, 10-15 Aug 2015, Rio de Janeiro, 
Brazil [C15-08-10.2]}



\author{Sara Bolognesi}
\email[]{sara.bolognesi@cea.fr}
\thanks{Speaker}
\affiliation{IRFU, CEA Saclay, Gif-sur-Yvette, France}


\date{December 2015}

\begin{abstract}
After highlighting the importance of neutrino cross section modeling for 
neutrino oscillation measurements, the most recent neutrino cross section measurements are presented.
New preliminary results are available from T2K for the measurement of charged current interactions 
on carbon without pions in the final state,
single pion production in water, coherent pion production in carbon and charged current inclusive
interactions in carbon as a function of neutrino energy. Few other results already published by the 
MINER$\nu$A and T2K collaborations are also discussed.  
\end{abstract}


\maketitle

\section{Neutrino cross section measurements}
Precise knowledge of the neutrino interaction cross section is crucial for present
and future long baseline neutrino oscillation experiments. The parameters describing neutrino 
oscillations are extracted by comparing the rate of neutrino interactions at near and far
detectors placed on the neutrino beamline. The near detector is sensitive to
the convolution of flux and neutrino cross section, this measurement is used to constrain the
neutrino spectrum expected at the far detector in absence of oscillations. In T2K the uncertainty on this measurement is
$\sim$8\% for $\nu$ and $\sim$11\% for $\bar{\nu}$, highly dominated by uncertainty on neutrino
cross section. In future long baseline experiments like DUNE~\cite{Adams:2013qkq} and HyperKamiokande~\cite{Abe:2011ts},
such uncertainty has to be kept below 2\% to avoid spoiling the sensitivity to CP-violation phase 
of the PMNS matrix of neutrino oscillation.

The extrapolation of the neutrino interaction rate from the near to the far detector is not straightforward
for various reasons:
\begin{itemize}
\item neutrinos at near and far detectors have different neutrino energy distributions
mainly because the neutrino spectrum, for a given neutrino flavor, is changed by the oscillation;
\item the near detector mainly measures $\nu_{\mu}$ and  $\bar{\nu}_{\mu}$, which dominate
the flux produced by the accelerator, while at the
far detector also $\nu_{e}$ and  $\bar{\nu}_{e}$, produced by oscillation, have to be measured;
\item near and far detectors have different acceptance for the outgoing particles produced 
in neutrino interactions;
\item near and far detectors may have different elemental composition and therefore
neutrino may interacts with different nuclear targets in the two detectors.
\end{itemize}
To perform the extrapolation from near to far detector,
the neutrino cross section needs thus to be known as a function of neutrino energy, for different neutrino
flavors and for anti-neutrinos, for different nuclear targets and
the distribution of the kinematics of the outgoing particles has to be known
(or, in other terms, exclusive cross sections computation are necessary).

On the other hand, the measurement of neutrino cross section is experimentally complicated since
the neutrino energy is not known event by event. The neutrino energy can be inferred
from the kinematics of the particles produced in the interaction but such approach is limited by
the detector precision: low energy particles can be reconstructed only above a given threshold,
the angular acceptance may be limited, the recoiling nucleus is mostly undetected and neutrons
are typically not detectable.
As a consequence there are large model uncertainties which are introduced in the unfolding of detector effects
to compute the signal efficiency and to estimate backgrounds. Best practices to address these issues include
quoting cross section measurements only in limited phase space with large and constant detector efficiency,
cross-checking results between different selections and analysis strategies and using control regions to 
constrain the backgrounds from data.

Finally, the produced neutrino energy spectrum and rate is known through the flux modeling, which is based on a detailed simulation
of the beam-line and, possibly, constrained by external hadro-production measurements. The flux uncertainties are typically
the largest systematics ($\sim$10\%) on the cross section overall normalization. To avoid such large uncertainties, cross section
ratios (between different nuclear targets, different neutrino species) can be measured. 

In the following, the most recent cross section measurements, at the time of NuFact15 conference, are reviewed.

\section{Quasi-Elastic-like interactions}
The Charged Current Quasi-Elastic (CCQE) interaction is the dominant one in the T2K neutrino energy spectrum, peaked around 0.6~GeV. 
In the events selected at the far detector for oscillation measurements, the neutrino energy
is computed from the angle and momentum of the outgoing lepton, assuming CCQE kinematics. Such energy reconstruction
from the lepton kinematics relies on assumptions on the nuclear model in the initial state (and on the distribution of the 
outgoing nucleon in the final state if this is below threshold). 
Since the oscillation parameters are extracted from the neutrino energy spectrum thus reconstructed, it is
crucial to have a very precise modeling of CCQE interactions to avoid biases on the measurement of
oscillation parameters.

The Monte Carlo simulations of the CCQE process rely on parametrizations that are tuned to data. 
In particular, the axial mass in the dipole form factor of the interaction ($M_A^{QE}$)
is tuned to old bubble-chamber data of neutrino interaction on deuterium. 
Additional nuclear effects in the nuclear targets, heavier than deuterium, used in modern experiments
(typically carbon, water or argon) are implemented relying on a Relativistic Fermi Gas
approximation to describe the nucleus (including corrections for Pauli blocking and binding energy). 
In 2010 MiniBooNE's measurement of CCQE cross section~\cite{AguilarArevalo:2010zc} has shown a large discrepancy with respect to this
simplified model. 
New models have been developed~\cite{Martini:2009,Martini:2010,Nieves:2012,Nieves:2012yz} which include long-range correlation between nucleons 
(computed in Random Phase Approximation, RPA) and neutrino interactions with correlated nucleon-nucleon 
pairs (called 2p2h). Such models have shown to describe successfully the MiniBooNE results. 
A good agreement with MiniBooNE measurements can also be obtained by tuning 
effective nuclear parameters (eg: $M_A^{QE}\sim$1.2 GeV)
but at the expense of disagreement with bubble-chamber data~\cite{Bernard:2001rs} 
(which give an axial mass $M_A^{QE}\sim$1~GeV).

The nuclear effects have to be taken into account not only in the initial state but also on the final state (FSI).
Neutrino interactions which produce a pion then absorbed in the nuclear medium by FSI cannot be distinguished 
experimentally from pure CCQE interactions.
The interactions measured experimentally are therefore
called Charged-Current Zero-Pions (CC0$\pi$) and consist of: pure CCQE interactions on single nucleon, 
interactions through the 2p2h channel (which includes $\Delta$ pion-less decay) and
single pion production from $\Delta$ resonance where the pion get absorbed by FSI effects (without affecting the muon kinematics).
It should be noted that the separation between FSI and $\Delta$ pion-less decay is somehow
arbitrary and can lead to double counting, a correct treatment (as in~\cite{Ankowski:2014yfa}) should consider the two contributions together
in the computation of the cross section.
The cited models including 2p2h are fully analytic and do not include FSI effects. 
On the other hand, long- and short-range correlations between nucleons have been recently
included in Monte Carlo generator like NEUT~\cite{Hayato:2002sd}, GENIE~\cite{Andreopoulos:2009rq} and NuWro~\cite{Juszczak:2009qa}, 
which include FSI effects through intranuclear cascade models.
Still these generators are tuned from data with effective parameters in order to get a satisfactory
agreement with the neutrino interaction measurements today available.

\begin{figure}
\begin{center}
 \includegraphics[width=5cm]{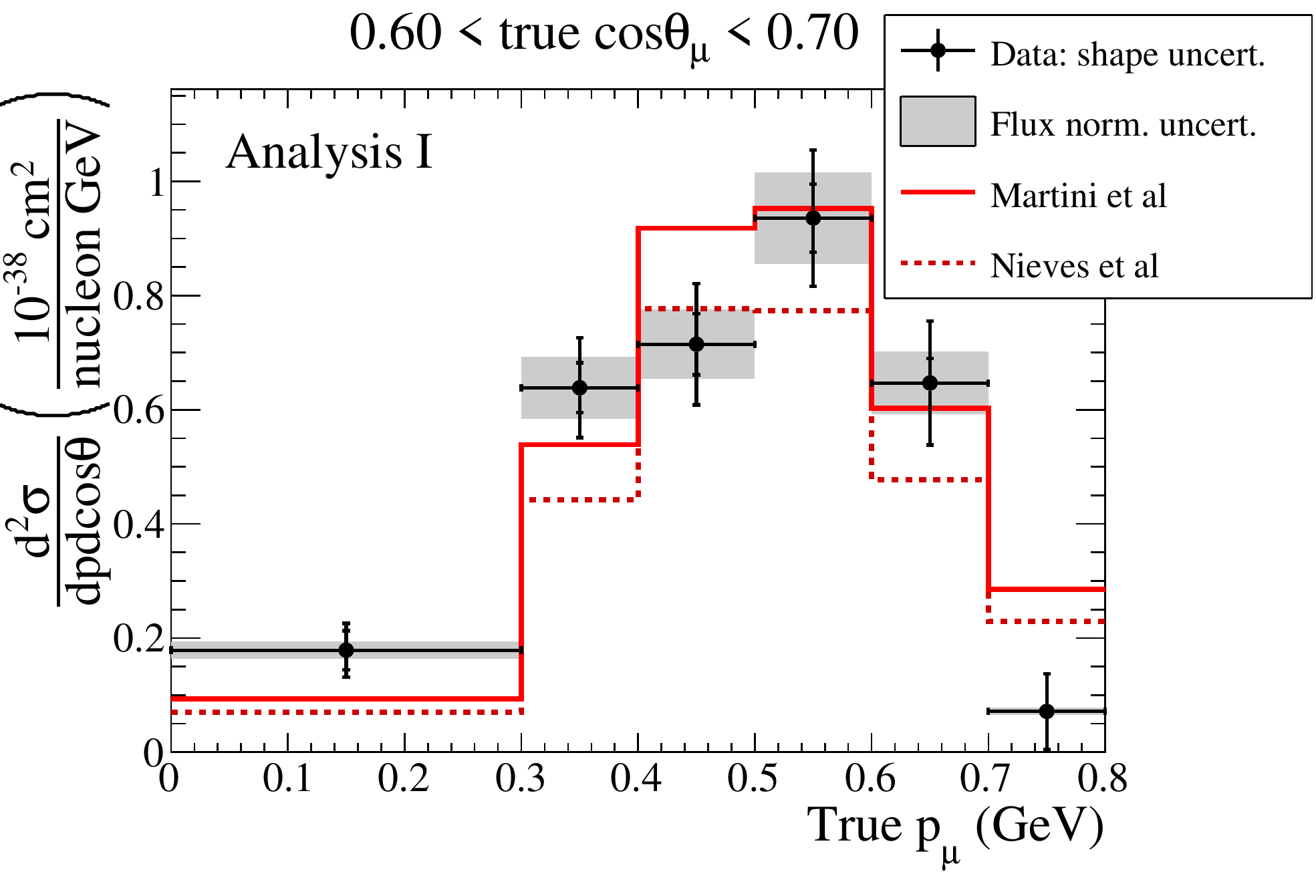}
 \includegraphics[width=5cm]{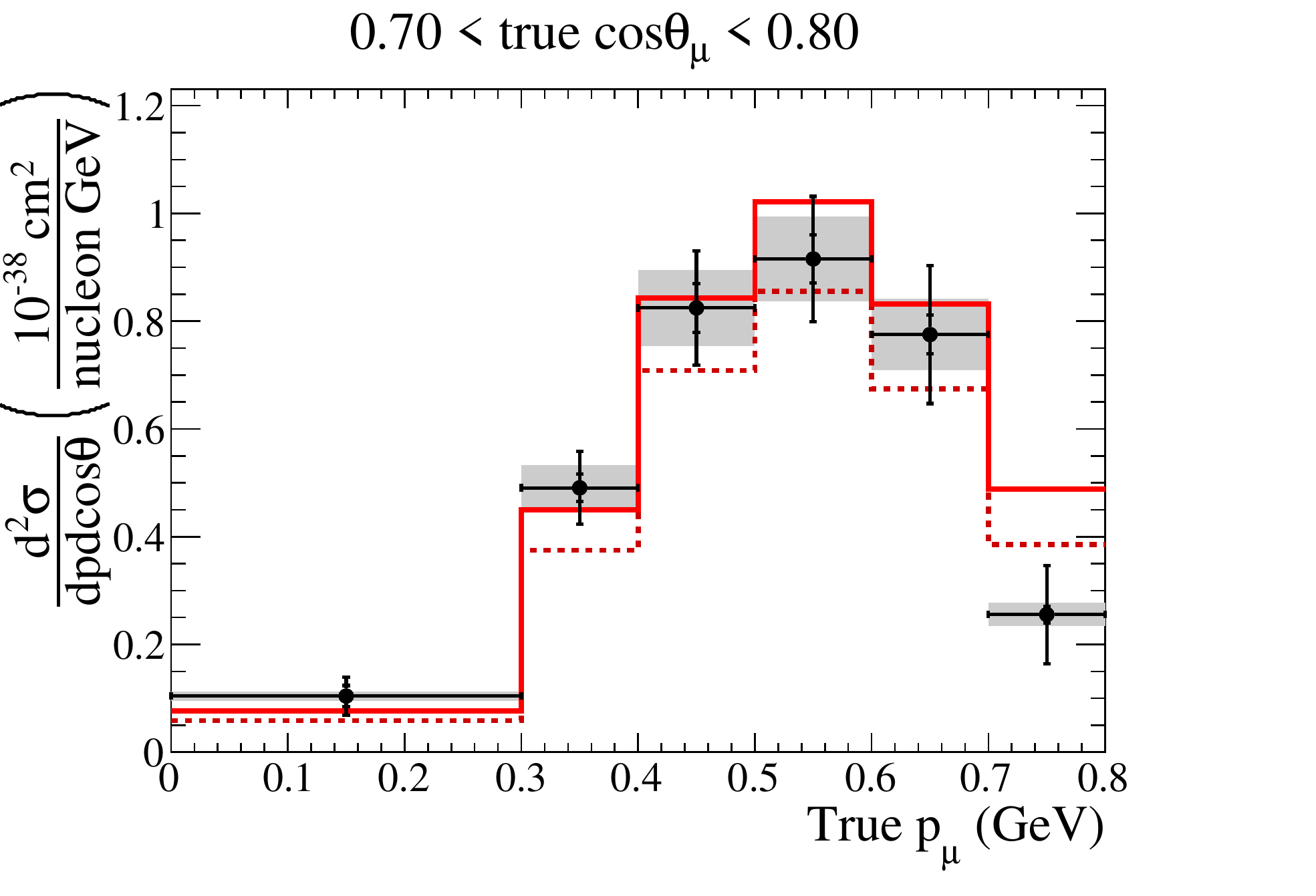}
 \includegraphics[width=5cm]{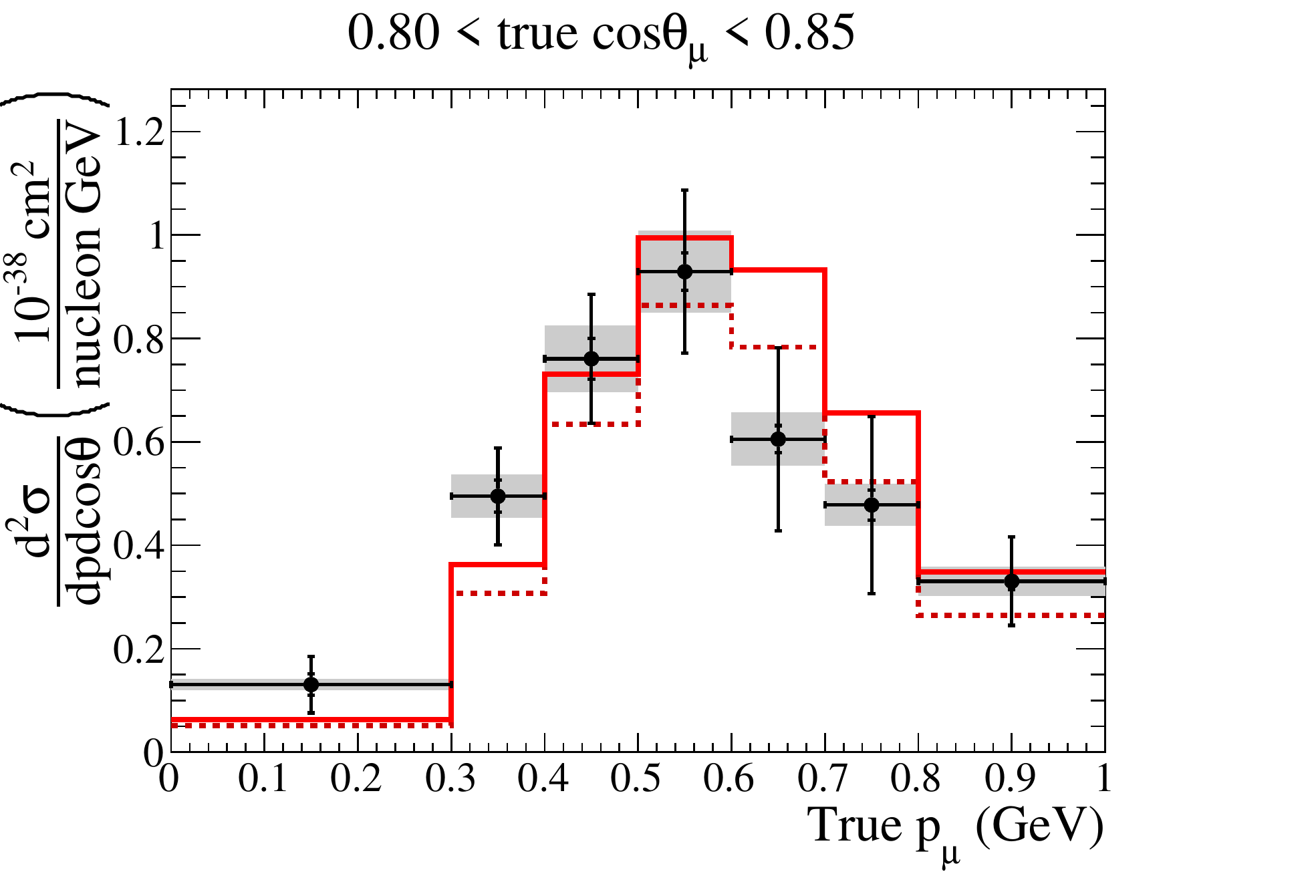}
 \includegraphics[width=5cm]{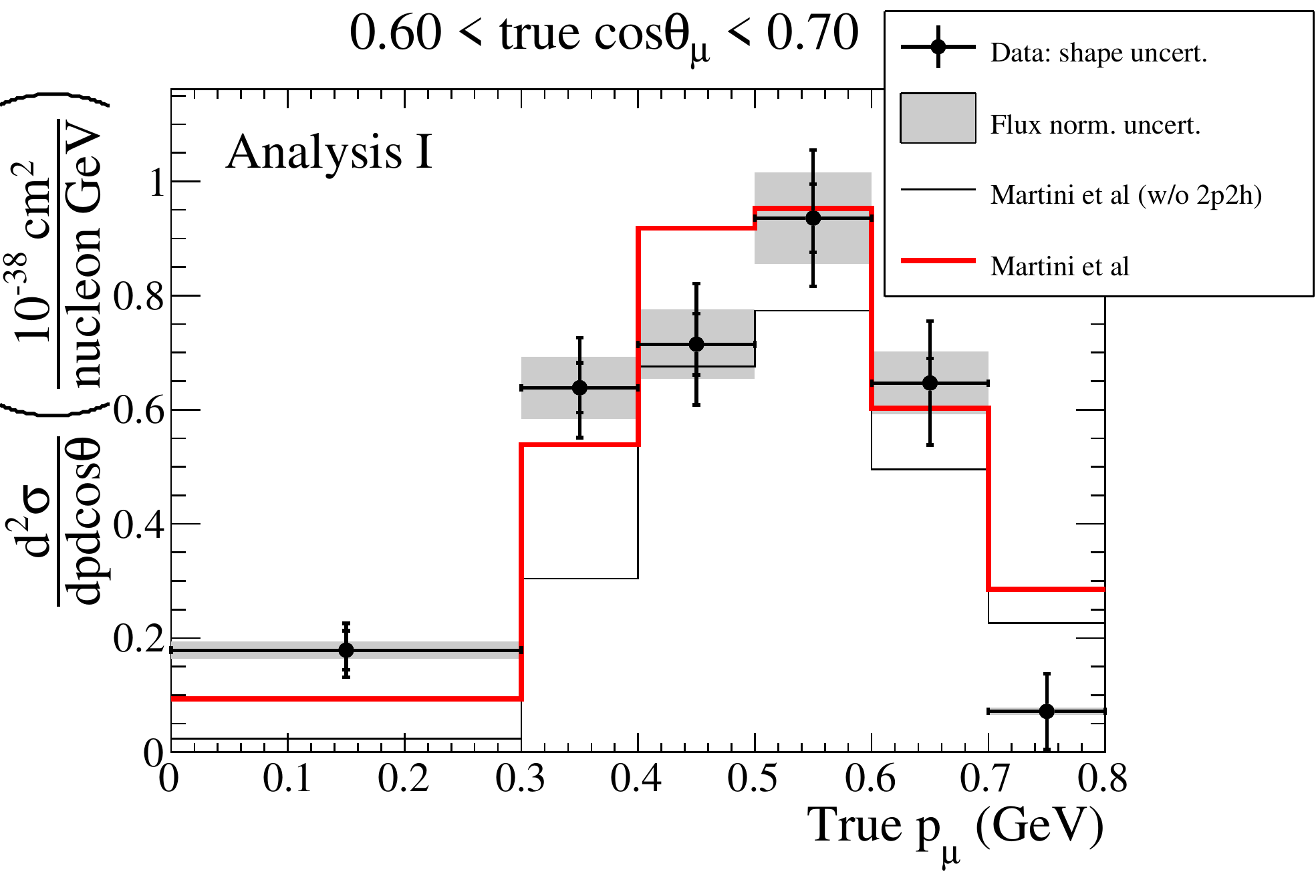}
 \includegraphics[width=5cm]{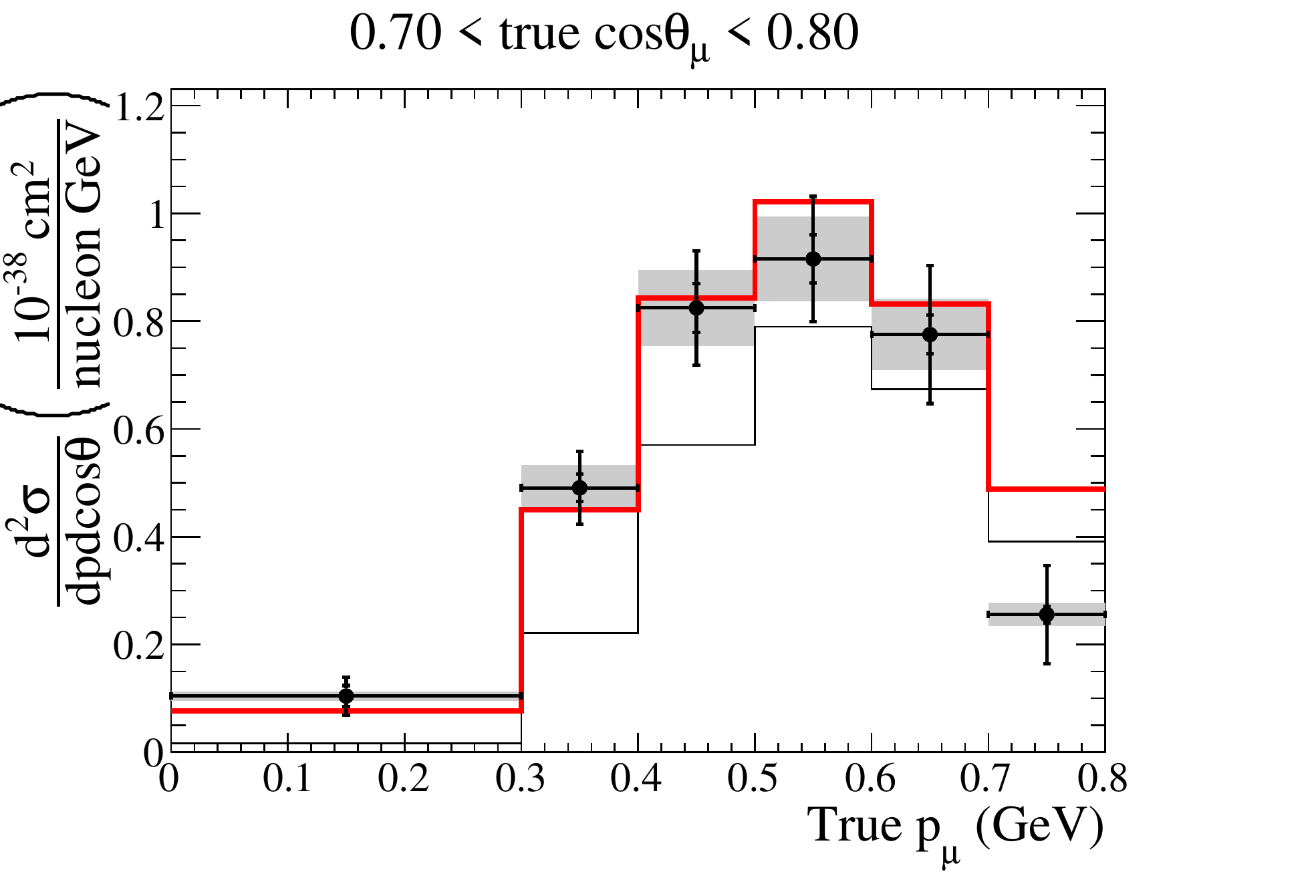}
 \includegraphics[width=5cm]{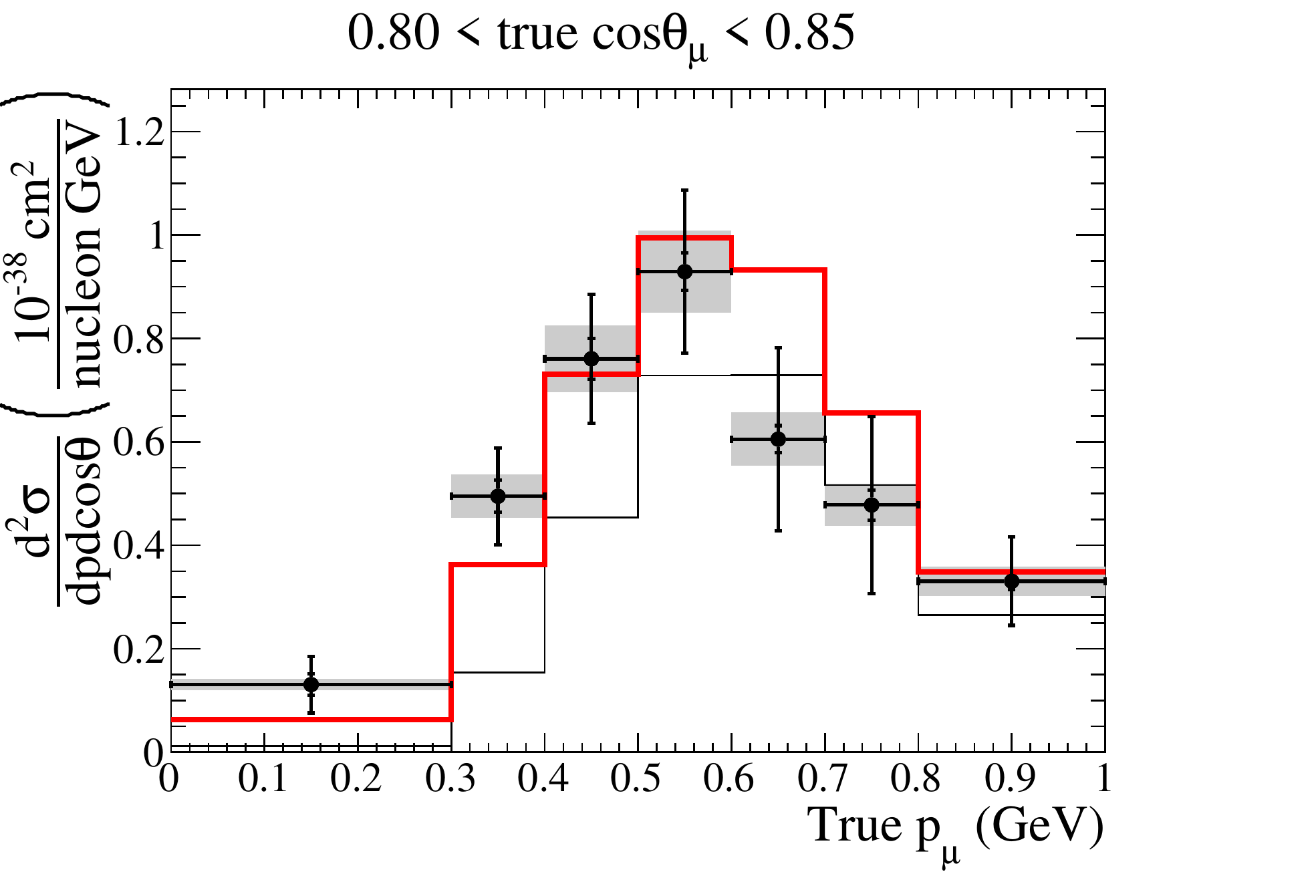}
 \includegraphics[width=5cm]{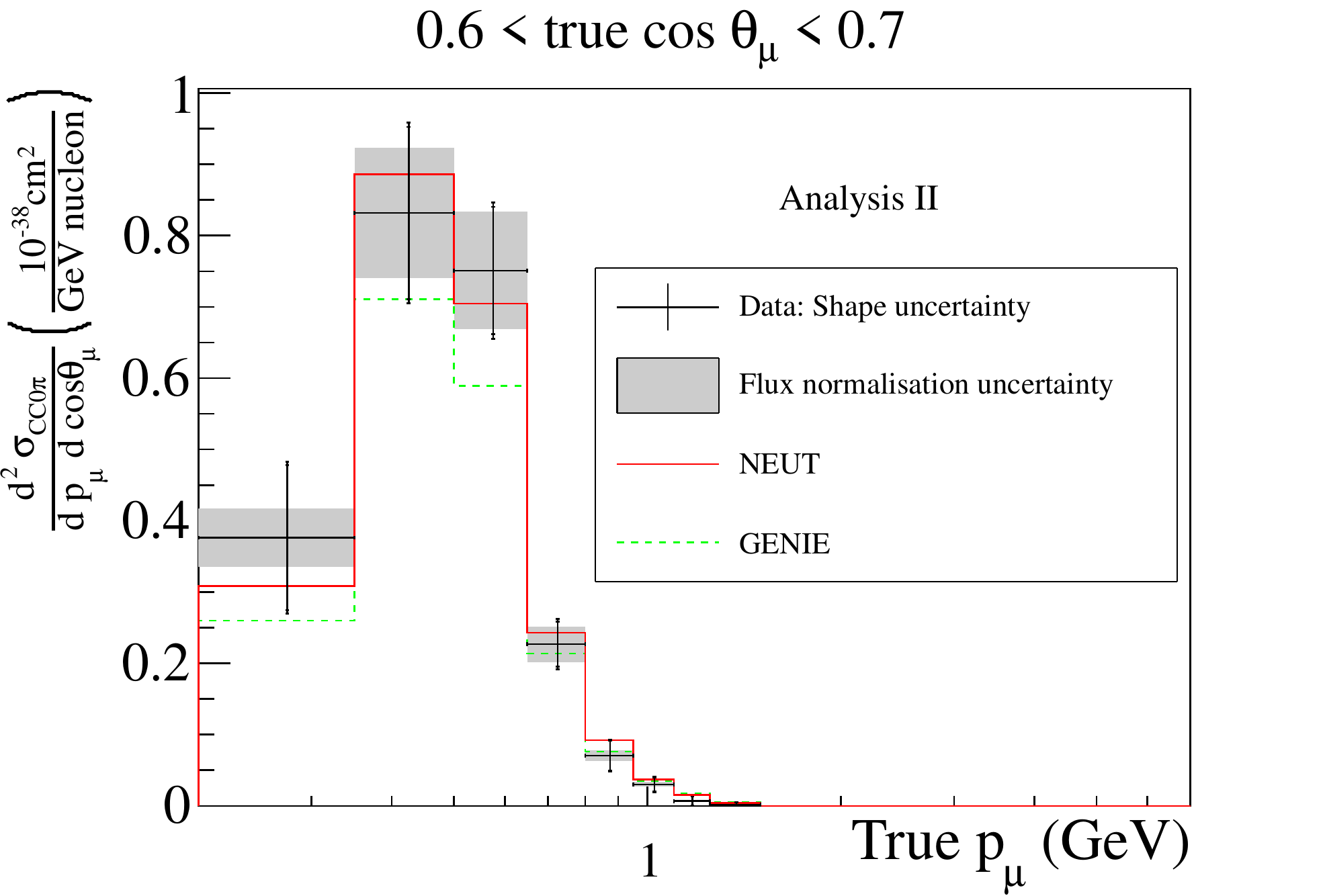}
 \includegraphics[width=5cm]{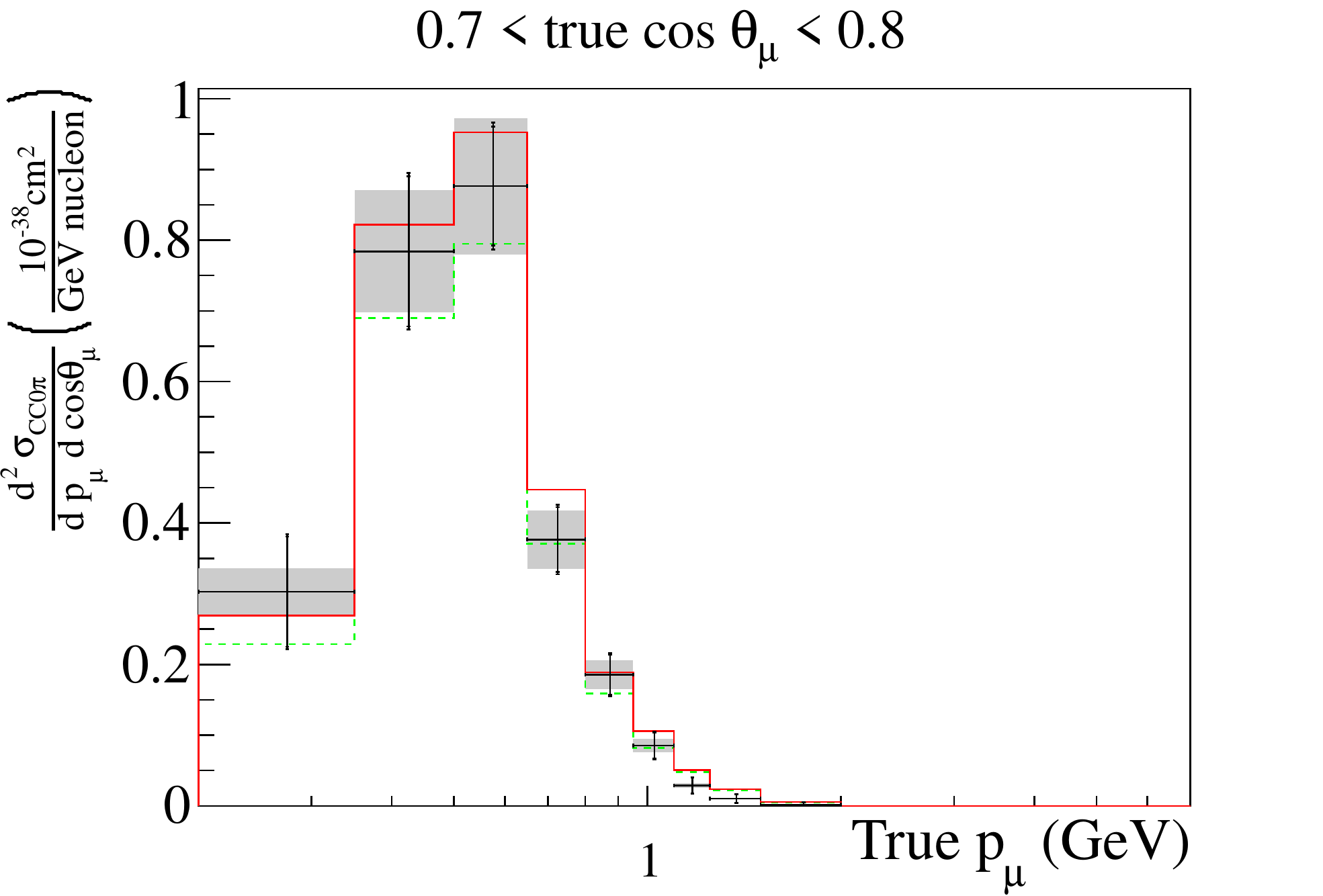}
 \includegraphics[width=5cm]{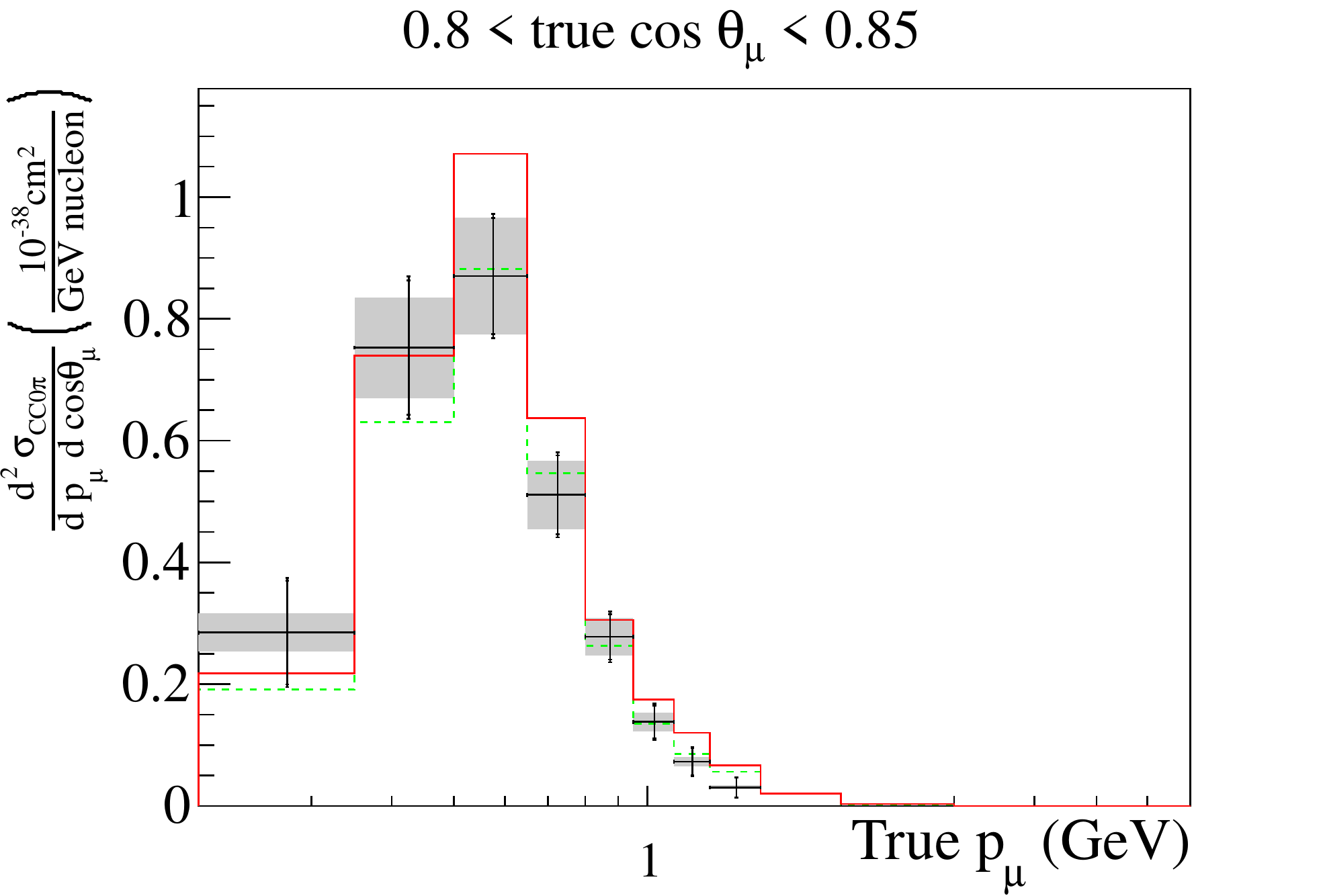}
\end{center}
\caption{Results of CC0$\pi$ measurement from T2K compared with the model of Martini et al and Nieves et al (top raw)
and with the model from Martini et al with and without 2p2h contribution (middle raw). 
Results of the same measurement with an alternative analysis strategy (bottom raw), compared to NEUT (v.5.1.4.2, $M_A^{QE}\sim$1.2 GeV) 
and GENIE (v2.6.4, $M_A^{QE}\sim$1.2 GeV) simulations without 2p2h.
\label{fig:CC0piT2K}}
\end{figure}

A new CC0$\pi$ measurement on carbon with the off-axis T2K near detector (ND280) is available. 
The analysis has been designed to be solid against model-dependent assumptions.
The selection requires events with only one reconstructed muon or a muon and a proton, special care has been taken to 
increase the efficiency to high angle and low momentum muons, the background prediction is tuned using control regions and the
result is presented as flux-integrated double-differential cross section as a function of muon momentum and angle.
A second analysis, based on different selection and cross section extraction method, has also been performed.
The agreement between the results from the two analyses proves the robustness of the measurement against the effects 
due to signal and background modeling. 
The results are compared in Fig.\ref{fig:CC0piT2K} to the predictions from Martini et al~\cite{Martini:2009,Martini:2010} 
and Nieves et al~\cite{Nieves:2012,Nieves:2012yz}, with and
without including multi-nucleons effects. Even if these models do not include FSI effects, the impact of CC1$\pi$ events with pion 
absorption is very small (few \%) in the intermediate angular and momentum region shown in Fig.\ref{fig:CC0piT2K}.
The results are also compared to Monte Carlo simulation including FSI effects but without nucleon-nucleon correlations.
The data prefer the presence of 2p2h contribution with respect pure CCQE with RPA corrections but the precision is not good enough yet to 
distinguish between different models.

\begin{figure}
\begin{center}
 \includegraphics[width=8cm]{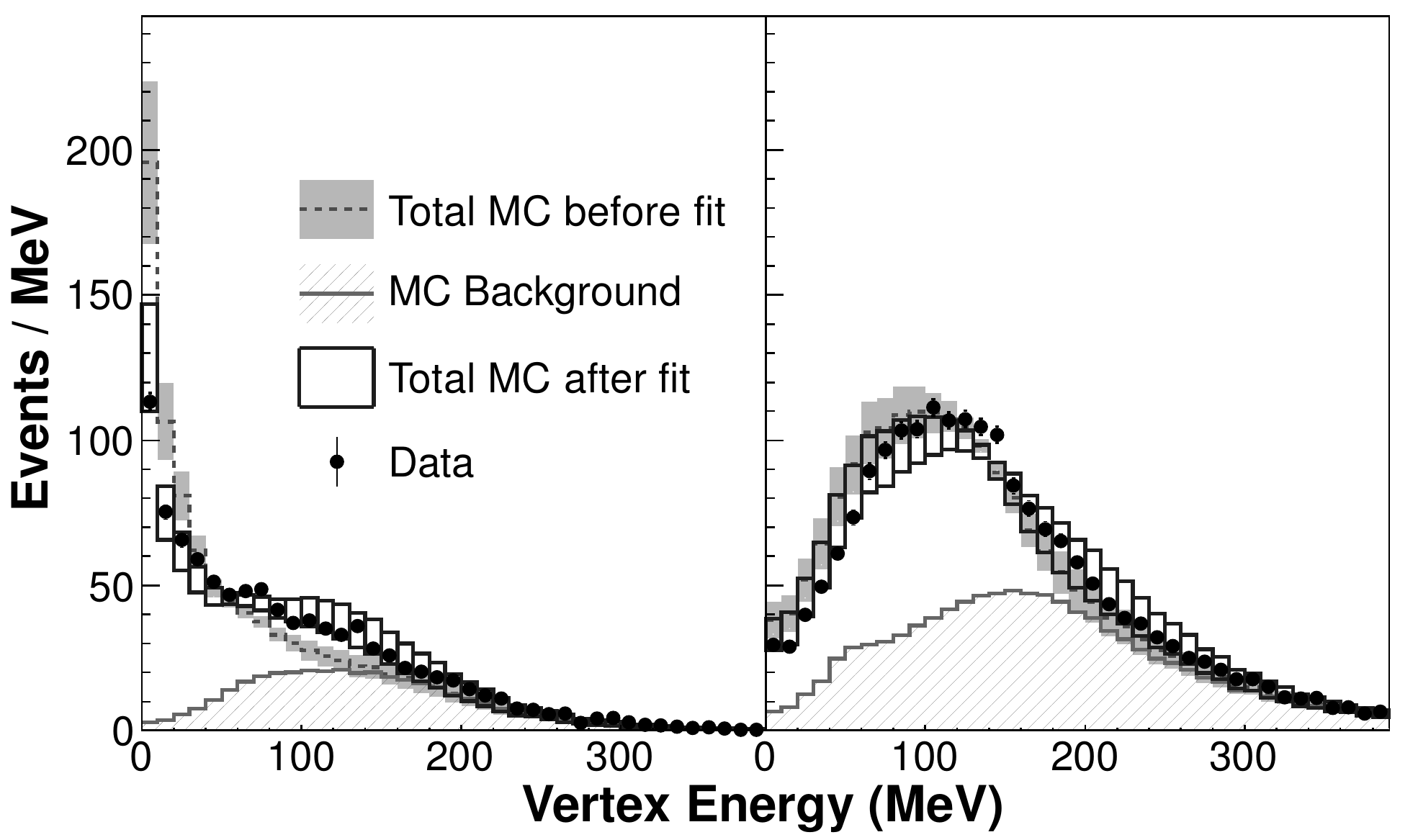}
\end{center}
\caption{Comparison between MINER$\nu$A data and simulation for the distribution of energy around the interaction vertex in CCQE events for
neutrinos (left) and anti-neutrinos (right).
\label{fig:vtxEneMinerva}}
\end{figure}

Measuring also the proton(s) angle and momentum may enhance the capability of distinguishing between different models. On the other hand
exclusive measurements are more model dependent, as the MINER$\nu$A analysis in~\cite{Walton:2014esl} and the T2K on-axis INGRID 
analysis in~\cite{Abe:2015oar}. 
To overcome such problem, it is interesting to look into variables as much as possible near to the actual experimental 
measurement, with minimal corrections for detector effects and acceptance. This is the case of the vertex energy distribution
published by MINER$\nu$A in~\cite{Fiorentini:2013ezn,Fields:2013zhk} and shown in
Fig.\ref{fig:vtxEneMinerva}. Considering that MINER$\nu$A detector is sensitive to proton but not to neutrons,
the excess at high vertex energy in neutrino data but not in anti-neutrino data, is an indication for the presence of
2p2h events ($\nu_{\mu} np\rightarrow \mu^- pp$, while  $\bar{\nu}_{\mu} np\rightarrow \mu^+ nn$ are undetectable). 
Unfortunately this kind of variables can only be compared with
Monte Carlo prediction fully embedded with detector simulation and the available models have today very poor predictive
power for the kinematics of the outgoing proton. The most informative and still model-independent measurement would be a cross section 
fully differential both in muon and proton angle and momentum, limited to the phase space of high proton and muon
reconstruction efficiency.

\section{Pion production}
\begin{figure}
\begin{center}
 \includegraphics[width=5.5cm]{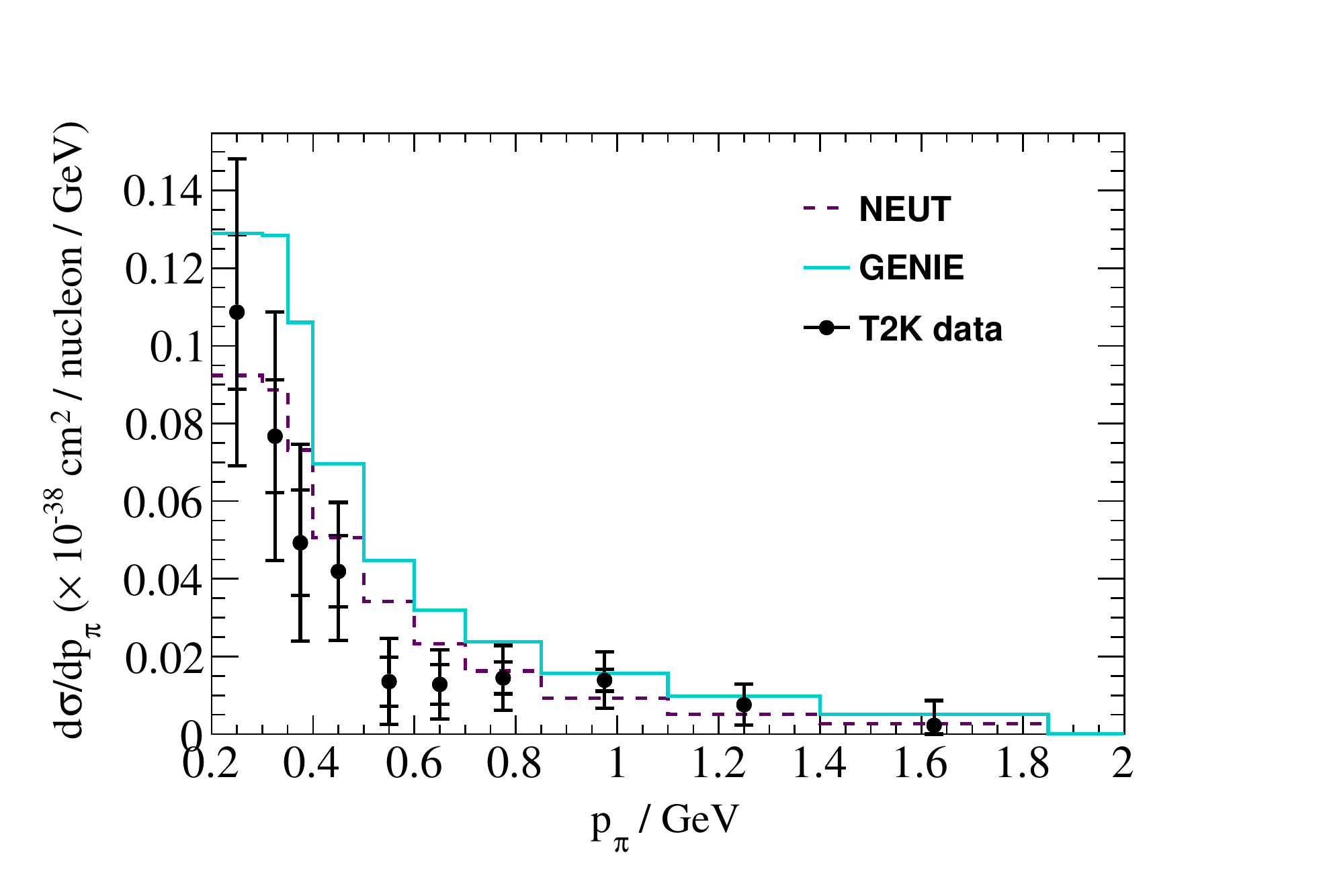}
 \includegraphics[width=5.5cm]{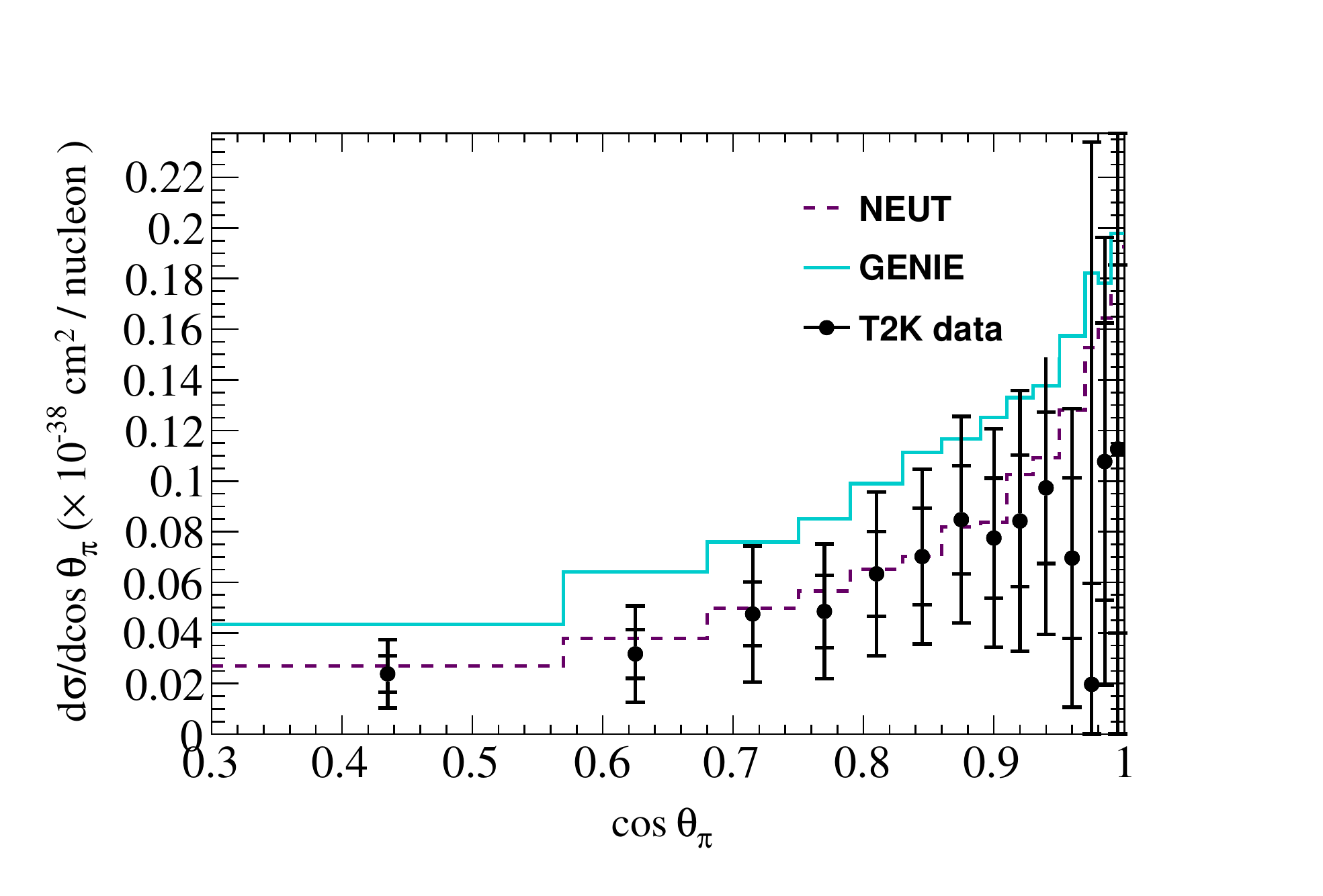}
 \includegraphics[width=5cm]{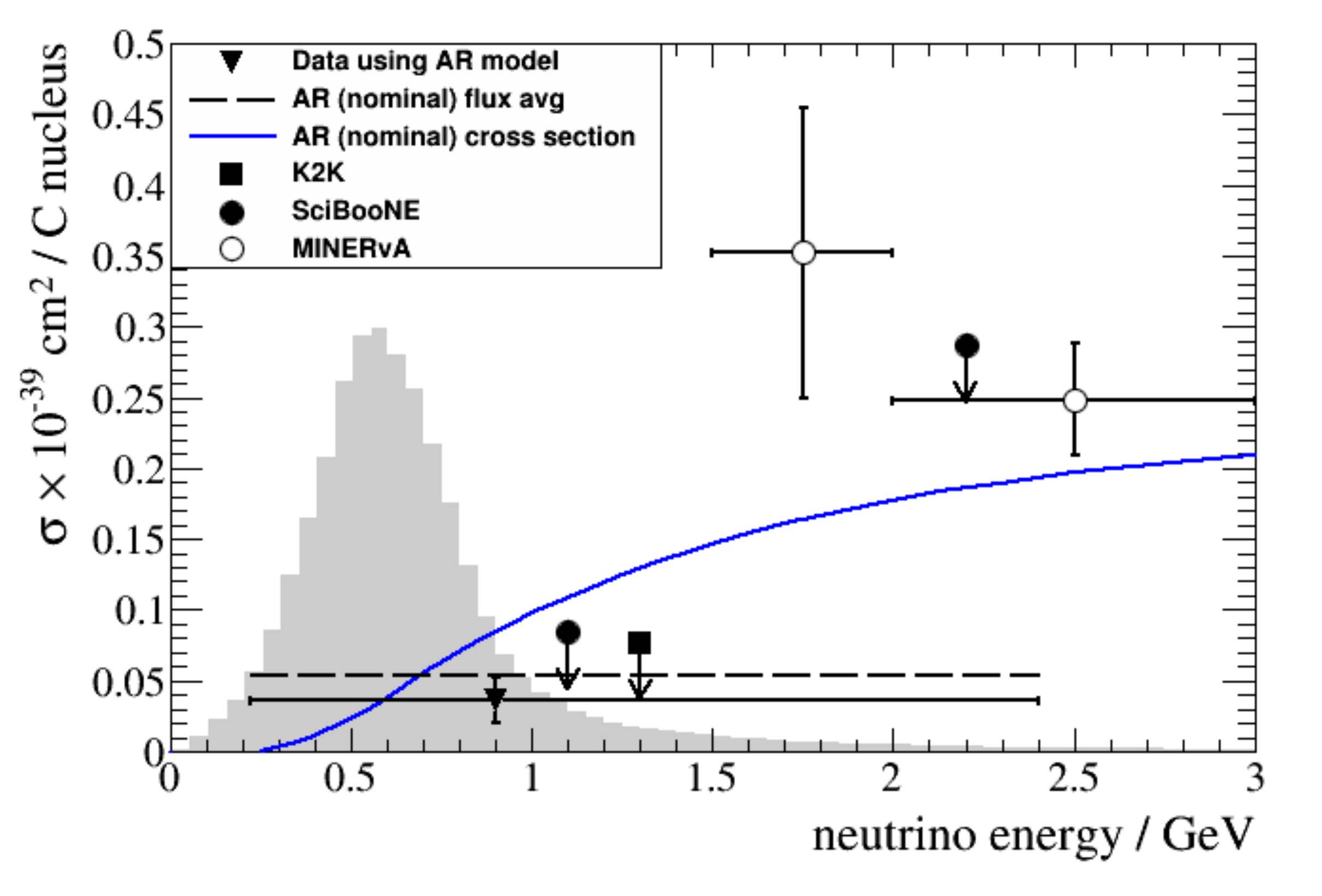}
\end{center}
\caption{T2K preliminary measurement of CC1$\pi^+$ cross section as a function of pion momentum (left) and angle (middle),
compared to NEUT and GENIE simulation. T2K measurement of coherent pion production (right) compared with expectation from Alvarez-Ruso model.
\label{fig:CC1piT2K}}
\end{figure}

The production of single pion in neutrino interaction is mainly due to $\Delta$ resonance production and decay.
FSI may then modify the kinematics of the pion, absorb it, change its charge, 
and/or produce other pions. It is well known that MiniBooNE~\cite{AguilarArevalo:2010bm} and MINER$\nu$A~\cite{Eberly:2014mra} results are in disagreement:
beyond overall normalization issues, the differences between the two experiments in the shape of the differential cross section as a function
of the outgoing pion energy cannot be described by any model (see, for instance,~\cite{Sobczyk:2014xza}). 
Since interaction cross section and FSI have different dependence on the number of nucleons, 
the two contributions can be disentangled by measuring CC1$\pi$ on different targets. 
T2K has new preliminary results for CC1$\pi^+$ on water with ND280 data. 
The signal includes events with only one pion and with positive charge. 
The target detector is composed of passive layers of water alternating
with two active layers of scintillator (CH). In order to be reconstructed, the outgoing pion in water interactions must reach
the first downstream active layer, therefore the acceptance is limited to relatively high pion momentum ($>200$~MeV).
In the first carbon active layer both carbon and water interactions are reconstructed
while the interactions reconstructed in the second layer are mainly due to carbon. Interactions on the second carbon layer are used 
to constraint the contribution of CC1$\pi$ on carbon on the first layer and thus extract the cross section due to water interactions only.
A control sample is used to constrain the background with multiple pions.
The results are shown in  Fig.\ref{fig:CC1piT2K}: GENIE tends to overestimate the overall rate while NEUT is in good agreement with the data; given the
present uncertainties, the shape is well reproduced by both generators, even if a hint of suppression is visible for very forward
pions. This region is dominated by interactions where the neutrino scatters coherently from an entire nucleus, leaving the nucleus unchanged
in its ground state. These interactions are characterized by very small momentum transferred to the nucleus $|t|=\sqrt{(q-p_\pi)^2}$ and
no hadronic activity (no nucleons is ejected). Such channel has been investigated recently by ArgoNeuT, MINER$\nu$A and T2K.
The reconstructed $|t|$ and the deposited energy around the vertex are used to select coherent interactions in data.
Sidebands in both $|t|$ and vertex energy distributions are used to tune the background.
It is indeed difficult to have a precise simulation of the background in the selected phase-space
because the vertex energy depends on the details of the detector response to very low energy deposits (near detector threshold and below)
and the model of non-coherent CC1$\pi$, especially for the selected kinematics, is not yet precisely constrained.
Previous results from MINER$\nu$A showed a suppression with respect to the Rein-Sehgal model~\cite{ReinSehgalCoh} at low pion energy and 
large pion angle, while ArgoNeuT statistics is not sufficient for a differential measurement but provide the first integrated
measurement on argon~\cite{Acciarri:2014eit}.
At low neutrino energy, where only upper limits (from K2K~\cite{Hasegawa:2005td} and SciBooNE~\cite{Hiraide:2008eu}) were available,
T2K has a new preliminary result (Fig.\ref{fig:CC1piT2K}): the coherent signal is observed with 2.2$\sigma$ significance and the results 
are in good agreement with a new microscopic model from Alvarez-Ruso~\cite{AlvarezRuso:2007tt}, better suited for low energy. 

\section{Dependence of the cross section on neutrino energy, nuclear target and neutrino flavor}
\begin{figure}
\begin{center}
 \includegraphics[width=4.5cm]{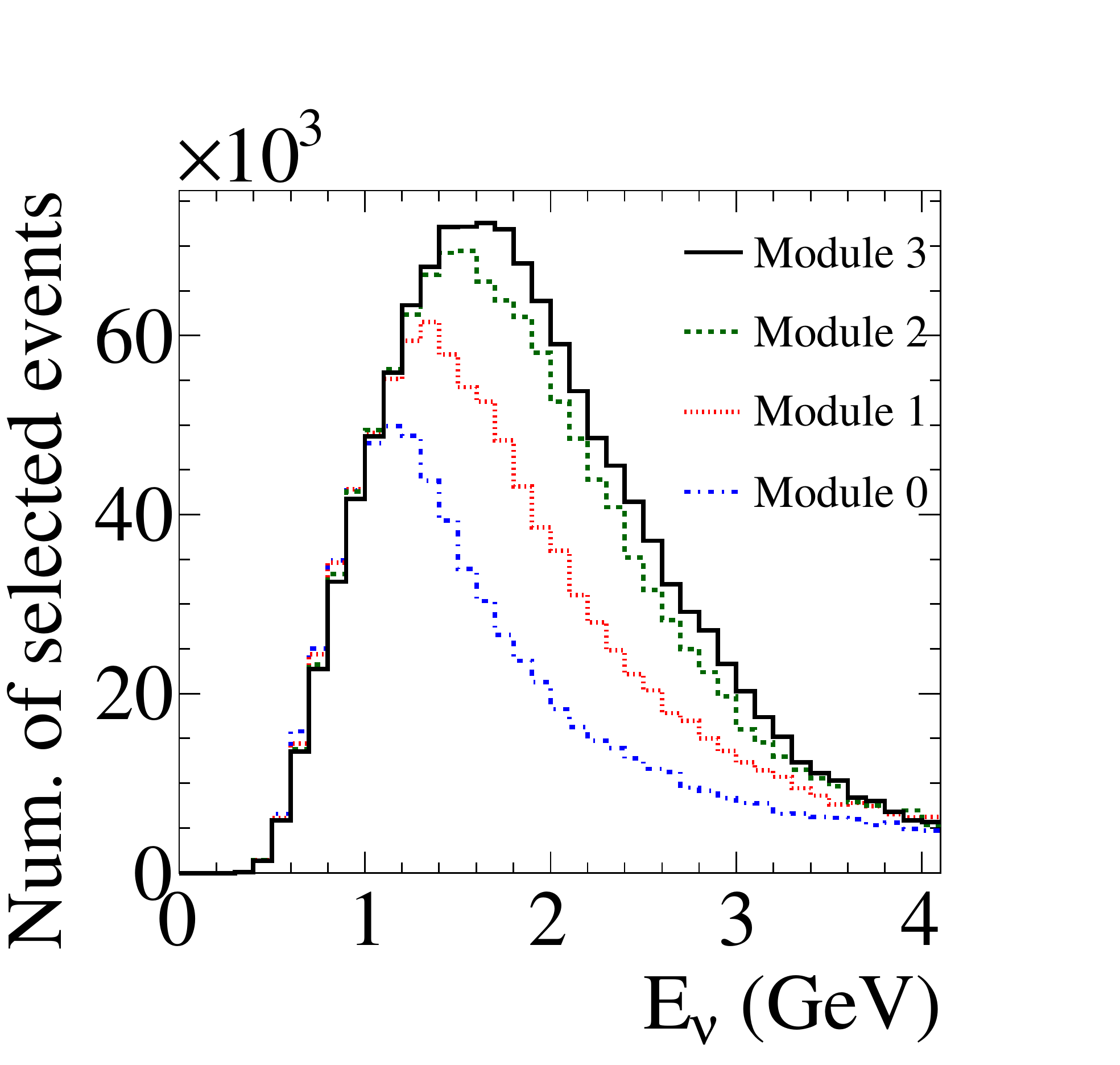}
 \includegraphics[width=6cm]{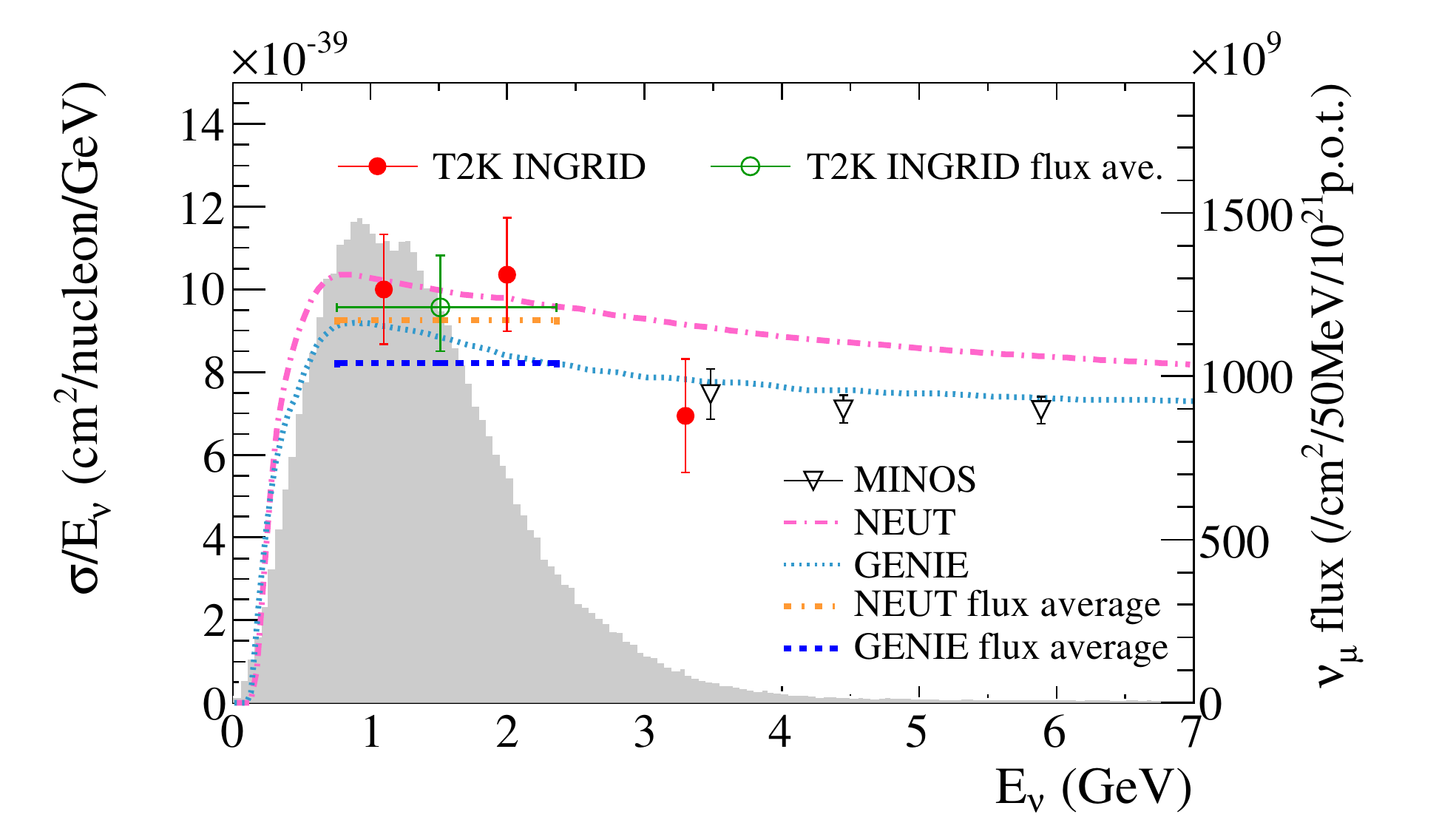}
 \includegraphics[width=4.5cm]{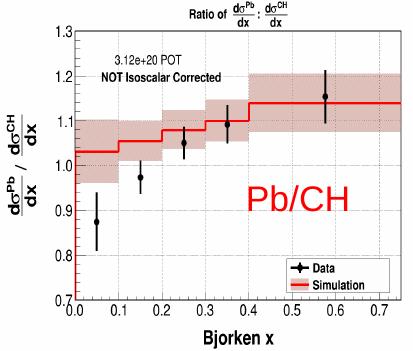}
\end{center}
\caption{Neutrino energy spectrum for events selected on different INGRID modules (left). CC-inclusive cross section measurement as a function of neutrino energy in INGRID (right). Ratio between lead and carbon DIS cross section as a function of Bjorken $x$ measured by MINER$\nu$A,
compared to GENIE simulation (right).
\label{fig:CCincINGRID}}
\end{figure}

As previously discussed, in the oscillation analysis to extrapolate the near detector constraints to the far detector, 
the energy dependence of the neutrino interaction cross section has to be known. Typically, the neutrino
energy can only be reconstructed from the kinematics of the outgoing particles relying on model-dependent
assumptions which introduce large theory systematics.
T2K has presented the first measurement~\cite{Abe:2015biq} which do not rely on such assumptions for the reconstruction of the neutrino energy. 
The idea is based on the geometry of the on-axis near detector: the INGRID detector is cross-shaped with the beam impinging 
in the center. Different modules are placed at different off-axis angles and therefore see different neutrino energy spectra, 
as shown in Fig.~\ref{fig:CCincINGRID}. The dependence on neutrino energy can then be extracted  by combining the neutrino interaction rate measured 
in different modules (the NuPRISM detector proposal~\cite{Bhadra:2014oma} is based on the same concept). 
In this approach the uncertainties in the flux modeling become the dominant systematics.
Results are shown in Fig.\ref{fig:CCincINGRID}: data suggest a suppression of the cross section at high energy.

If the near and the far detector have different elemental composition (as in T2K), it is crucial to know 
the cross section dependence on the nuclear target. Comparing the cross section for different nuclear targets is also
useful to isolate and measure nuclear effects. The ratio between cross sections on different targets has the advantage of
canceling the flux uncertainties as well as most of the systematics related with model-dependent assumptions
on the signal efficiency and background estimation. To maximize such cancellation, it is crucial
to impose the same phase space for the particles outgoing from interactions on both the nuclear targets.
This has been done in the T2K measurement of the CC-inclusive, per nucleon, cross section ratio between iron and carbon 
in INGRID~\cite{Abe:2014nox}:
\begin{equation}
\frac{\sigma^{Fe}_{CC}}{\sigma^{CH}_{CC}}=1.047\pm 0.007(stat.)\pm 0.035(syst.),
\end{equation}
where the acceptance of outgoing muons for events both on carbon and on iron, has been limited with kinematics cut only in the region accessible
for carbon events (small muon angle). The total cross-sections are then corrected to the full phase space but the theory systematics 
due to this correction mostly cancel out in the ratio.

An interesting measurement of ratio between targets has been presented by MINER$\nu$A 
in the region dominated by Deep Inelastic Scattering (DIS)~\cite{Tice:2014pgu}. 
A recent update, after background subtraction and correction for detector effects, of lead over carbon cross section ratio as a function
of Bjorken $x$, shown in Fig.\ref{fig:CCincINGRID}, suggests slightly larger nuclear screening effects for low $x$ than what
is in the GENIE simulation. 
In these measurements on different nuclear targets, as for the CC1$\pi$ measurement
in water by T2K previously discussed, the interactions may happen on carbon in the scintillator region or in
relatively thin layers of passive material, made of different nuclear targets, surrounded by scintillator. 
In order to reconstruct the interaction, the outgoing particles must exit from the 
passive layer and reach the downstream scintillator module. Extrapolating two or more tracks to their common
starting point, the event vertex may be assigned to the passive layers but the procedure is limited by the tracking precision.
There is also the possibility that backward going particles leak into
the scintillator region upstream of the passive layer which may confuse the vertex reconstruction algorithm.
As a consequence, there could be migration of events: interactions which have the real vertex in the passive layer
may be reconstructed as scintillator interactions and viceversa. This is the main systematics common to
the available measurements on nuclear targets which are not active detectors. Since the modeling of backward nucleons 
is not well known theoretically, the control of this event migration must be done
from data using control regions of interactions in the scintillator. 

\begin{figure}
\begin{center}
 \includegraphics[width=7cm]{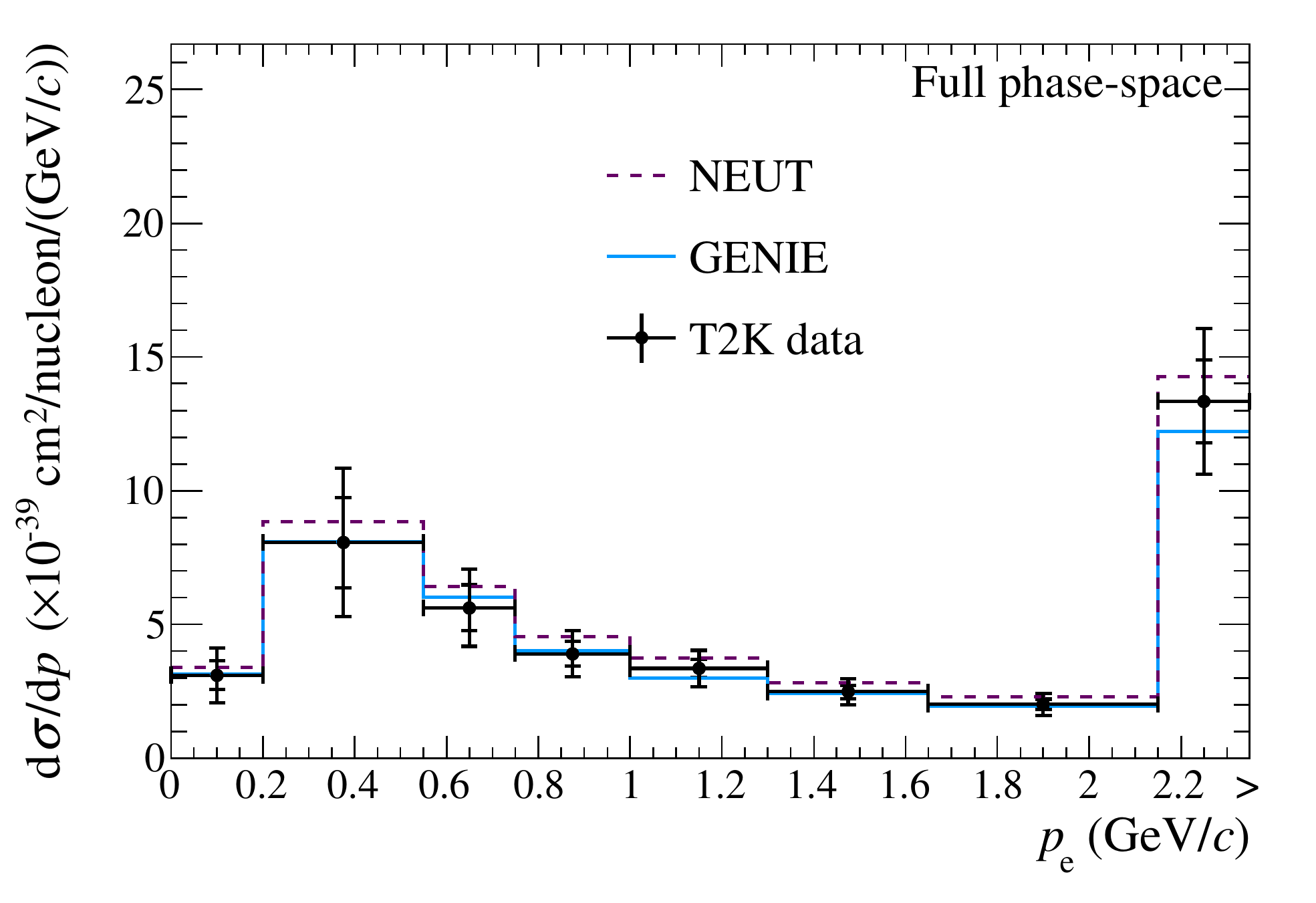}
 \includegraphics[width=7cm]{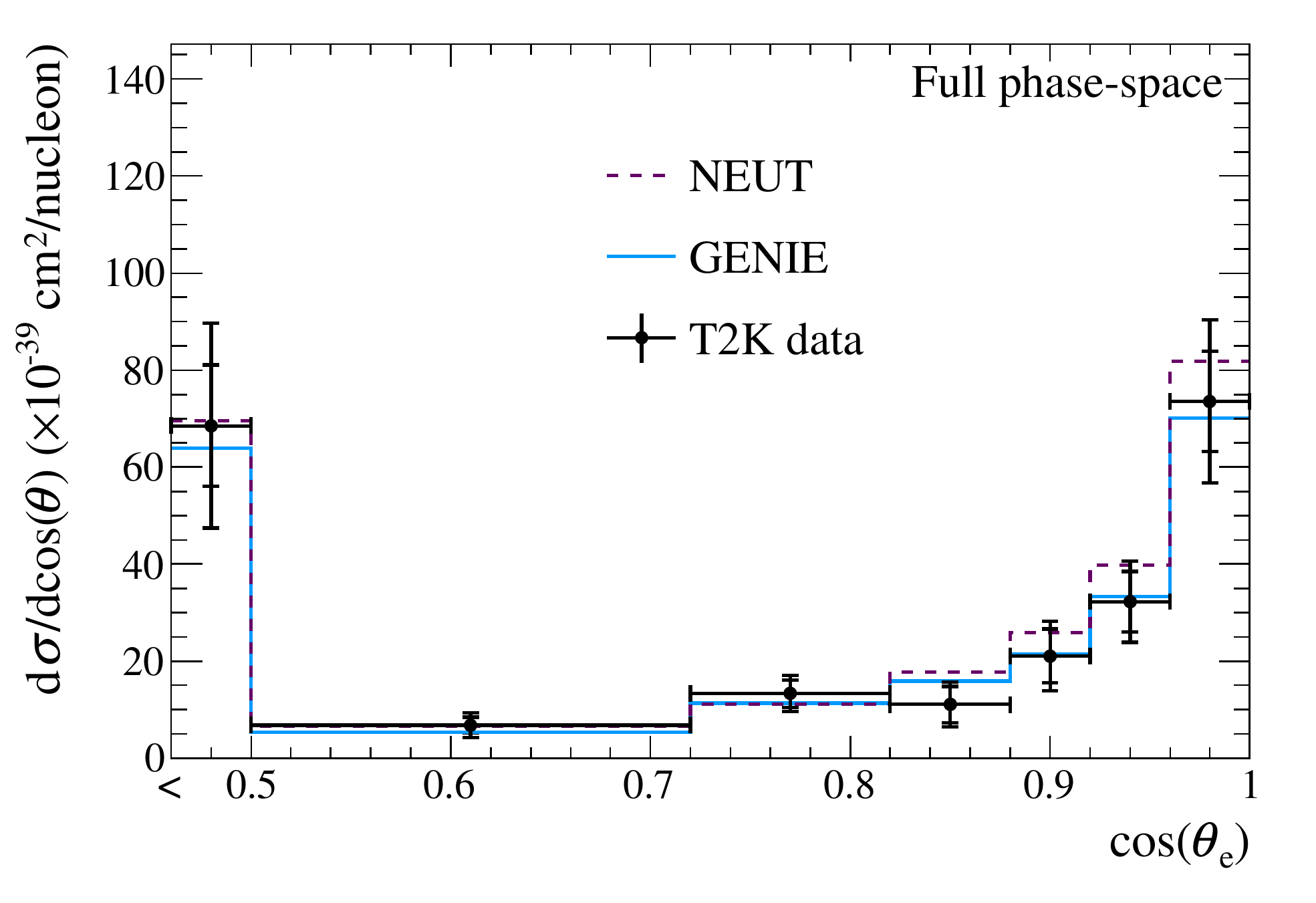}
\end{center}
\caption{Measurement of CC-inclusive electron neutrino cross section on carbon at T2K as a function of electron momentum (left) and angle (right). \label{fig:nuE}}
\end{figure}

Finally for future long baseline experiments, which will dispose at the far detector of large statistic samples of $\nu_e$ appearance,
is particularly important to measure the cross section for electron neutrino. At the near detector the flux
of $\nu_e$ is very small with respect to $\nu_\mu$ ($\Phi_{\nu_e}/\Phi_{\nu_\mu}<$1\%), so this measurements are mainly limited by statistics.
T2K has measured the CC-inclusive cross section on carbon as a function of the electron momentum and angle~\cite{Abe:2014agb}, as shown in Fig.~\ref{fig:nuE}. 
T2K has also measured the ratio between data and NEUT simulation of the total rate of electron neutrino CC-inclusive interactions 
in water~\cite{Abe:2015mxf}:
\begin{equation}
R_{on-water}=0.87 \pm 0.33 (stat.) \pm 0.21 (syst.).
\end{equation}

\section{Conclusion}
For long baseline neutrino oscillation measurements it is crucial to know the neutrino-nucleus
interaction cross section with high precision in order to perform an unbiased extrapolation of the near detector
constraints to the far detector. Considering the accuracy expected at future experiments (DUNE, Hyper-Kamiokande), the
present uncertainty on the neutrino cross sections would be the limiting systematics.

For future neutrino cross section measurements, in view of the poor knowledge of the nuclear effects involved 
in the neutrino-nucleus interactions,
it is particular important to design analyses as much as possible solid against model-dependent
assumptions. The cross section measurements are always affected by initial and final state
interactions which are difficult to disentangle. In order to reach a detailed understanding of the different
effects, it is crucial to compare measurements from different interaction processes, at different neutrino energies,
on different nuclear targets and for different neutrino species. On the experimental point of view, it is important
to compare results from various experiments which are limited by different systematics and to cross-check results
from different neutrino fluxes, the flux uncertainty being the dominant systematics on the cross section normalization.

Due to the complexity of the problem and in view of the importance for future long baseline experiments, 
it is fundamental to maintain a long term effort for neutrino interactions measurements based on a strict collaboration 
between different experiments and with the theory community.

\bibliography{NuFact15template}

\end{document}